\documentclass[preprint]{aastex}
\usepackage{graphicx}
\usepackage{natbib}
\usepackage{comment}
\defcitealias{dawson11}{Paper I}

\begin{document}

\title{Molecular Clouds in Supershells: A Case Study of Three Objects in the walls of GSH 287+04--17 and GSH 277+00+36}
\author{J. R. Dawson\altaffilmark{1}}
\email{joanne.dawson@utas.edu.au}
\author{N. M. McClure-Griffiths\altaffilmark{2}}
\author{John M. Dickey\altaffilmark{1}}
\author{Y. Fukui\altaffilmark{3}}
\altaffiltext{1}{School of Mathematics and Physics, University of Tasmania, Sandy Bay Campus, Churchill Avenue, Sandy Bay, TAS 7005}
\altaffiltext{2}{Australia Telescope National Facility, CSIRO Astronomy \& Space Science, Marsfield NSW 2122, Australia}
\altaffiltext{3}{Department of Physics and Astrophysics, Nagoya University, Chikusa-ku, Nagoya, Japan}

\begin{abstract}
We present an in-depth case study of three molecular clouds associated with the walls of 
the Galactic supershells GSH 287+04--17 and GSH 277+00+36. These clouds have been identified in previous work as examples in which molecular gas is either being formed or destroyed due to the influence of the shells. $^{12}$CO(J=1--0), $^{13}$CO(J=1--0) and C$^{18}$O(J=1--0) mapping observations with the Mopra telescope provide detailed information on the distribution and properties of the molecular gas, enabling an improved discussion of its relationship to the wider environment in which it resides. We find that massive star formation is occurring in molecular gas likely formed in-situ in the shell wall, at a Galactic altitude of $\sim200$ pc. This second-generation star formation activity is dominating its local environment; driving the expansion of a small H{\sc ii} region which is blistering out of the atomic shell wall. We also find new morphological evidence of disruption in two smaller entrained molecular clouds thought to pre-date the shells. We suggest that at the present post-interaction epoch, the lifetime of this surviving molecular material is no longer strongly determined by the shells themselves.
\end{abstract}


\section{Introduction}

The last decade has seen the emergence of a new paradigm for molecular cloud formation, in which the production of star-forming molecular gas is integrated into the modern picture of a dynamic, turbulent interstellar medium (ISM). In this picture, the compression, cooling and fragmentation of the diffuse atomic medium in colliding turbulent flows produces clumpy sheets and filaments of cold material, which may go on to become molecular and self-gravitating if density and column density conditions are met \citep{hennebelle99,koyama00,hartmann01,bergin04,audit05,vazquez06,vazquez07,heitsch08c,heitsch08b,inoue09}. 

Within the scope of physical systems covered by this theory are supershells -- large loops, shells and cavities, often hundreds of parsecs in diameter, that are formed by the cumulative stellar feedback from OB clusters \citep[e.g.][]{heiles79,mccray87,mcclure02}. Multiple stellar winds and supernovae drive a shock front into the surrounding medium, which persists over large enough size-scales and timescales that the transition from atomic to molecular gas may occur in the cooled material of the swept-up shell.

Previous work by \citet[][hereafter Paper I]{dawson11} has presented parsec-resolution observations of the atomic and molecular ISM in two Galactic supershells, in which we find an enhanced level of molecularization over the volumes of both objects. This provides the first direct observational evidence of increased molecular cloud production due to the influence of supershells, demonstrating that in the case of these two objects the global influence on the molecular ISM is a positive one. However, our observations also highlight another long-recognized fact -- that the influence of shells may also be disruptive to molecular gas on local scales; driving the atomic-molecular transition in the opposite direction in pre-existing clouds disrupted by the shell's passage. 	

Here we zoom in on three objects from this previous work in order to present a detailed case study of molecular clouds in supershell walls. We carry out 
observations in the $^{12}$CO(J=1--0), $^{13}$CO(J=1--0), and C$^{18}$O(J=1--0) lines with the Mopra telescope, and combine these with H{\sc i} data from \citetalias{dawson11} and \citet{mcclure03}, to examine objects in which the molecular-atomic transition is likely proceeding in opposite directions. The Mopra data reveal the molecular gas at a resolution of $\sim45\arcsec$, providing detailed information on its distribution and properties, and enabling a greatly improved discussion of its relationship to the wider environment in which it resides.

For the remainder of this introduction we describe the target objects, giving background details on the properties of the molecular clouds and their parent shells. The CO observing and data reduction strategy are described in Section \ref{observations}. Section \ref{results} deals in turn with the three clouds; presenting molecular gas distributions, deriving masses and column densities, assessing star formation activity, and presenting an in-depth examination of each in terms of its relationship to its local environment and parent shell. Our findings are summarized in Section \ref{summary}. 

\subsection{Target Objects}

\begin{figure}
\epsscale{1.0}
\plotone{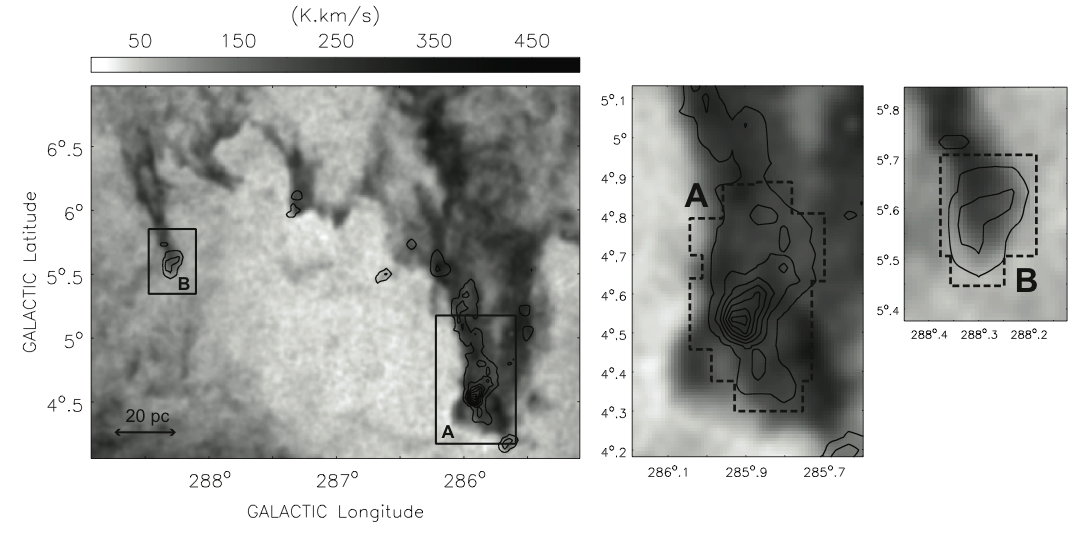}
\caption{Integrated intensity image of a subsection of the wall of GSH 287+04--17. The greyscale image is H{\sc i} observed with the ATCA and Parkes telescopes, and contours are $^{12}$CO(J=1-0) observed with the NANTEN telescope \citep[both lifted from][]{dawson11}. The velocity integration range is $-26.5 < v_{lsr} < -19.9$ km s$^{-1}$, and CO contour levels are drawn at 1.5+5.0 K km s$^{-1}$. Dashed lines show regions mapped with the Mopra telescope in the present work.}
\label{fig:287regions}
\end{figure}

\begin{figure}
\epsscale{1.0}
\plotone{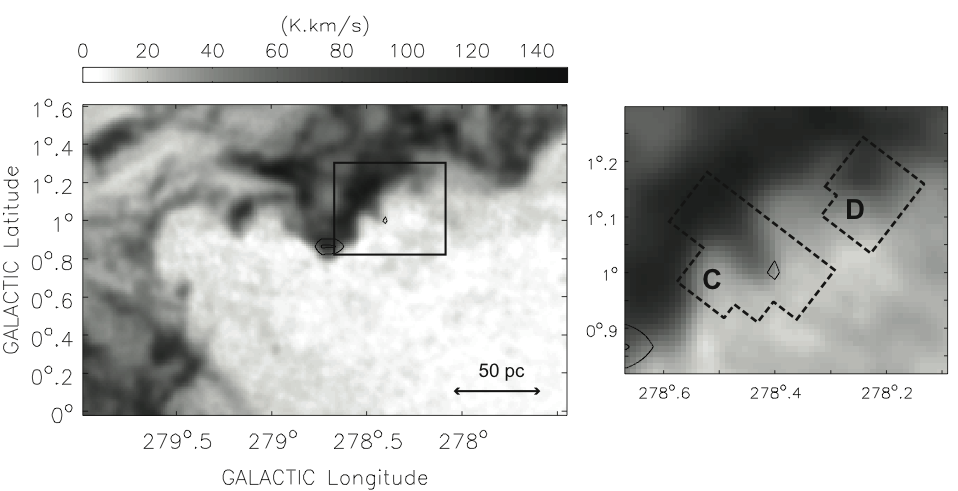}
\caption{Integrated intensity image of a subsection of the wall of GSH 277+00+36. The greyscale image is H{\sc i} observed with the ATCA and Parkes telescopes, and contours are $^{12}$CO(J=1-0) observed with the NANTEN telescope \citep[both lifted from][]{dawson11}. The velocity integration range is $41.6 < v_{lsr} < 44.1$ km s$^{-1}$, and CO contour levels are drawn at 1.2+1.0 K km s$^{-1}$. Dashed lines show regions mapped with the Mopra telescope in the present work.}
\label{fig:277regions}
\end{figure}

We observe three known molecular clouds in the Galactic supershells GSH 287+04--17 and GSH 277+00+36. Both shells are large, evolved objects, with radii of $\gtrsim150$ pc, swept-up masses of $\gtrsim10^6M_{\odot}$, ages of $\gtrsim10^7$ yr, and gentle expansion velocities of $10$--$20$ km s$^{-1}$ \citep[for details see][]{fukui99, mcclure00, mcclure03, dawson08a, dawson11}. Their distances are estimated as $2.6\pm0.4$ kpc and $6.5\pm0.9$ kpc respectively, with GSH 287+04--17 located in the inner and GSH 277+00+36 in the outer Galaxy. Crucially, both contain large quantities of associated molecular gas, equivalent to $\sim10$--$20\%$ of their total neutral gas mass. The target regions are shown in Figures \ref{fig:287regions} and \ref{fig:277regions}. An additional fourth region covers a small section of shell wall in which no CO has been previously detected. 

Cloud A comprises the lower half of a large ($M\sim7700~M_\odot$) molecular cloud embedded in the wall of GSH 287+04-17. At a distance of $z\sim200$ pc above the Galactic midplane, its altitude is several times greater than the local molecular gas scale height \citep[$\sigma_z\approx60$ pc, e.g.][]{malhotra94}. This object was first noted by \citet{fukui99} and later appears in the cloud catalogue of \citet{dawson08b} as G285.90+4.53 (cloud 16), detected in both $^{12}$CO(J=1--0) and $^{13}$CO(J=1--0). \citetalias{dawson11} identifies this cloud as a likely example of molecular material formed in-situ in the shell walls, based on the configuration of the H{\sc i} and CO emission, the consistency of formation timescales, and the presence of sufficient material for shielding of the CO molecule. The molecular material forms an integral and coherent part of the shell wall, with CO emitting gas distributed along the curved rim of the atomic shell, and a tight correlation between the H{\sc i} and CO emission observed throughout the feature. A bright, young star forming region dominates the lower regions of the cloud. 

Cloud B is also located in the wall of GSH 287+04--17, and is identified as G288.27+5.60 (cloud 92) in the catalogue of \citet{dawson08b}. The shell walls show tapered fingers of H{\sc i} pointing radially inwards towards the shell interior, some of which harbor clumps of molecular gas at their tips. With an estimated mass of $M\sim900~M_\odot$, Cloud B is the largest of these molecular clumps, and is located at the very tip of an extended tail of H{\sc i}. Unlike Cloud A, this object is not associated with large H{\sc i} column densities or well-embedded within a section of wall. It is instead offset towards the very edge of the H{\sc i} finger, with CO emission extending outside the detected limits of the atomic feature. \citetalias{dawson11} identifies this cloud as a likely example of remnant molecular material that existed in the region before the shell's passage, which has subsequently 
become disrupted and entrained in the shell interior. Morphological signatures and a lack of material for shielding of the CO molecule attest to this interpretation. This cloud is also a high altitude object; located at a distance of $z\sim250$ pc from the Galactic midplane. 

Cloud C is a small molecular cloud located at the tip of an H{\sc i} `drip'  in the wall of GSH 277+00+36. This cloud was detected at a single position in $^{12}$CO(J=1--0) in \citetalias{dawson11} and has not yet been mapped. It's location at the tip of an H{\sc i} feature marks it as another candidate for pre-existing molecular material that is interacting with the expanding shell. Like Clouds A and B, it is also located some distance from the Galactic midplane, at $z\sim170$ pc.

The region labelled `D' in Figure \ref{fig:277regions} covers another H{\sc i} `drip' in the walls of GSH 277+00+36, in which no molecular emission has previously been detected. Its appearance and location are both similar to the feature that hosts Cloud C.

\section{Observations}
\label{observations}

The four target regions were observed with the Mopra telescope, near Coonabarabran, Australia.  Observations of the GSH 277+00+36 regions were made in June 2005 and were targeted towards positions of high H{\sc i} intensity, since no molecular detections had previously been published at that time. Observations of the GSH 287+04-17 clouds were made in October 2006 and October 2009, and covered all positions at which $^{12}$CO(J=1-0) emission had been previously detected with the NANTEN telescope \citep{fukui99}.

The transitions observed were the $^{12}$CO(J=1--0), $^{13}$CO(J=1--0) and C$^{18}$O(J=1--0) lines, at 115.3, 110.2 and 109.8 GHz. In 2005 the Mopra telescope was equipped with a dual band SIS receiver that provided 600 MHz instantaneous bandwidth between 86 and 115 GHz. For $^{12}$CO observations the correlator was configured to provide 1024 channels over a bandwidth of 64 MHz, providing a velocity resolution of $0.16$ km s$^{-1}$ and a velocity coverage $\sim165$ km s$^{-1}$. For $^{13}$CO and C$^{18}$O the spectral resolution was doubled to 0.08 km s$^{-1}$, reducing the total velocity coverage to $\sim80$ km s$^{-1}$. The 2006 and 2009 sessions made use of the newly installed MMIC receiver and the MOPS digital filter bank, which can simultaneously record dual polarization data for up to sixteen 137.5 MHz zoom bands positioned within an 8 GHz window. Zoom bands centered on the rest frequencies of the three lines each contained 4096 channels, providing a velocity resolution and coverage of 0.09 km s$^{-1}$ and $\sim360$ km s$^{-1}$ in all lines. The remaining zoom bands were positioned to cover other dense gas tracers, but no emission was detected.

Observations were made in an on-the-fly (OTF) raster mapping mode, in which the telescope records data continuously while scanning across the sky. Mosaics of $5'\times5'$ maps were arranged to cover the target regions, with an overlap of 12\arcsec\ between each map to ensure smooth coverage. Each map was observed at least twice in orthogonal scanning directions to minimize scanning artifacts. For the 2005 session the scan speed was 3 arcsec\ s$^{-1}$ and spectral data was recorded at 6\arcsec\ intervals along the scan direction. The spacing between scan rows was 8\arcsec\ in $^{12}$CO and 9\arcsec\ in $^{13}$CO and C$^{18}$O. In 2006 and 2009 the scan speed was 3.5 arcsec\ s$^{-1}$, the sampling interval was 14\arcsec\ and the spacing between scan rows was 10\arcsec. These settings fulfill the minimum requirements for oversampling of the $\sim33\arcsec$ (FWHM) Mopra beam. For all sessions an off-source position was observed once per scan row. 

The pointing solution of the telescope was verified once every 60--100 minutes via observations of the SiO masers X\_Cen, VY\_CMa, RW\_Vel and R\_Car, and corrections were applied for pointing errors of greater than 10$''$ in either azimuth or elevation. System temperature measurements were made with a paddle at the start of each map and at 30 minute intervals thereafter, and took typical values of $450$--$650$ K at 115 GHz and $250$--$350$K at 109 and 110 GHz. 

Bandpass calibration, baseline subtraction and calibration onto a $T_A^*$ scale are performed with the \textit{livedata} package. The spectra are then gridded into cubes using \textit{gridzilla}.
\footnote{Binaries and source code for \textit{livedata} and \textit{gridzilla} are available from http://www.atnf.csiro.au/computing/software/livedata.html.} The data are weighted by the inverse of the system temperature, and convolved with a truncated Gaussian smoothing kernel with a FWHM of $60\arcsec$ and cutoff radius of $30\arcsec$ to improve signal-to-noise ratio. This results in a final effective angular resolution of $45\arcsec$. Post-gridding, a $27\arcsec$ band was trimmed from the outer edge of each image to remove noisy map edges.   

The gridded data are then converted to a main beam temperature scale using a scaling factor determined by daily observations of the standard calibrator source Orion KL ($\alpha_{B1950}=5^h32^m47.5^s$, $\delta_{B1950}=-5^{\circ}24\arcmin21\arcsec$). \citet{ladd05} have shown that for $^{12}$CO(J=1--0) and $^{13}$CO(J=1--0) Orion KL peak main beam temperatures recorded with Mopra match those measured by the 15m SEST telescope, demonstrating the source is sufficiently extended within the beams of both instruments that measurements from the two may be directly compared. Referring to SEST documentation, and assuming a SEST main beam efficiency of $\eta_{\mathrm{mb}}=0.7$ \citep{johansson98}, scaling factors of $\eta=0.53\pm0.02$, $0.40\pm0.02$ and $0.53\pm0.02$ are required to rescale the Mopra $^{12}$CO(J=1--0) data for 2005, 2006 and 2009, respectively. The first and last of these figures are in excellent agreement with the documented Mopra `extended main beam efficiency', $\eta_{\mathrm{xb}}\approx0.55$, appropriate for observations of sources with angular extent $\gtrsim80''$ \citep{ladd05}. The outlying 2006 figure is consistent with data recorded by other Mopra observers in the same period \citep{hughes10}. The equivalent scaling factors derived for $^{13}$CO(J=1--0) are $0.59\pm0.02$ (2005), $0.59\pm0.02$ (2006) and $0.58\pm0.03$ (2009). These are also applied to the C$^{18}$O(J=1--0) line. 

In cases where a region is covered by data from two different years, the spectra are re-gridded to a common grid and weighted by the inverse of their rms noise before being linearly combined. Finally, narrow channel data are binned by two velocity channels, bringing the final velocity resolution of all datasets to 0.16--0.18 km s$^{-1}$. The rms noise per channel varies across the observed regions, and takes values of 300--600 mK in $^{12}$CO(J=1--0), and 100--200 mK in $^{13}$CO(J=1--0) and C$^{18}$O(J=1--0).

\section{Results \& Discussion}
\label{results}

$^{12}$CO(J=1--0) and $^{13}$CO(J=1--0) emission was detected in Clouds A, B and C, and C$^{18}$O(J=1--0) was detected in Clouds A and B only. No CO was detected in Region D, to a $3\sigma$ detection limit of 1.0 K in $^{12}$CO(J=1--0). 

\subsection{Derivation of Column Densities \& Masses}

Before examining each of the three clouds in depth, we briefly outline the common analysis performed for each. 

\begin{figure}
\epsscale{1.0}
\plotone{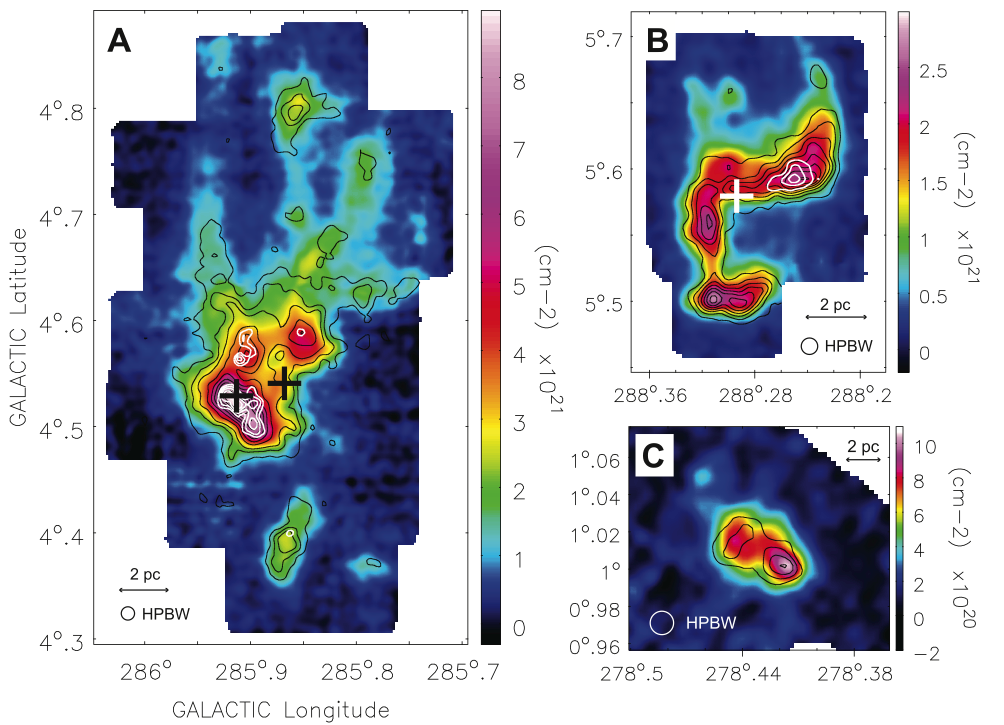}
\caption{$N_{\mathrm{H}_2}$ maps of Clouds A, B and C. Color images show material traced in $^{12}$CO(J=1--0), black contours in $^{13}$CO(J=1--0) and white contours in C$^{18}$O(J=1-0). Contour levels begin at the $3\sigma$ detection limit for each tracer and are as follows. Cloud A -- $^{13}$CO: [0.55, 1.05, 1.55, 2.55, 3.55, 4.55, 5.55, 6.55, 7.55, 8.55]$\times10^{21}$ cm$^{-2}$; C$^{18}$O: [4.5, 5.5, 6.5, 7.5]$\times10^{21}$ cm$^{-2}$. Cloud B -- $^{13}$CO: [0.40, 0.65, 0.9, 1.15, 1.40, 1.65, 1.9]$\times10^{21}$ cm$^{-2}$; C$^{18}$O: [2.5, 3.5, 4.5]$\times10^{21}$ cm$^{-2}$. Cloud C -- $^{13}$CO: [0.15, 0.25, 0.35, 0.45]$\times10^{21}$ cm$^{-2}$.
In cloud A, C$^{18}$O contours are only plotted within the $^{12}$CO(J=1--0) cloud boundary (i.e. noise peaks  from high-noise regions outside the cloud have been removed). Crosses show IRAS point sources with moderate or high quality flux measurements at 60 $\mu$m and 25 $\mu$m, and $F_{60} > F_{25}$.}
\label{fig:nh2all}
\end{figure}

Figure \ref{fig:nh2all} shows H$_2$ column density maps for the three clouds. For material traced in $^{12}$CO(J=1--0) we use an X-factor of $X = 1.6\times10^{20}$ K$^{-1}$ km s$^{-1}$ cm$^{-2}$ \citep{hunter97} to convert CO integrated intensity directly to molecular hydrogen column density, $N_{\mathrm{H}_2}$. %
This treatment implicitly assumes that the $^{12}$CO line is optically thick and that the molecular clouds are self-gravitating \citep[see e.g.][]{heyer01}. 
For $^{13}$CO(J=1--0) and C$^{18}$O(J=1--0), $N_{^{13}\mathrm{CO}}$ and $N_{\mathrm{C}^{18}\mathrm{O}}$ may be obtained from the following relations under the assumption of local thermodynamic equilibrium (LTE):\\
\begin{equation}
N_{^{13}\mathrm{CO}}/\mathrm{cm}^{-2}=~2.42\times10^{14}~\left[\frac{T_{ex}~\Delta v}{1-\exp(-5.29/T_{ex})}\right]~\sum_i\tau_{^{13}\mathrm{CO},i}~,
\end{equation}
\begin{equation}
N_{\mathrm{C}^{18}\mathrm{O}}/\mathrm{cm}^{-2}=~2.54\times10^{14}~\left[\frac{T_{ex}~\Delta v}{1-\exp(-5.27/T_{ex})}\right]~\sum_i\tau_{\mathrm{C}^{18}\mathrm{O},i}~,
\end{equation}\\
where $\Delta v$ is the channel width in km s$^{-1}$ and the sum is performed over all channels, $i$, within the cloud velocity range. Here, the optical depth in the $i$th channel, $\tau_i$, is computed as
\begin{equation}
\tau_{^{13}\mathrm{CO},i}= -\ln \left[1-\frac{T_{mb}{_{,^{13}\mathrm{CO},i}}}{5.29\left(\left\{\exp(5.29/T_{ex})-1\right\}^{-1}-0.164\right)}\right],
\end{equation}
\begin{equation}
\tau_{\mathrm{C}^{18}\mathrm{O},i}= -\ln \left[1-\frac{T_{mb,i}{_{,\mathrm{C}^{18}\mathrm{O},i}}}{5.27\left(\left\{\exp(5.27/T_{ex})-1\right\}^{-1}-0.166\right)}\right].
\end{equation}\\
CO column densities are then converted to $N_{\mathrm{H}_2}$ by assuming H$_2$-CO abundance ratios of $5\times10^5$ for $^{13}$CO \citep{dickman78} and $6\times10^6$ for C$^{18}$O \citep{frerking82}.

In the above expressions, the excitation temperature, $T_{ex}$, is estimated from the peak temperature of the $^{12}$CO(J=1--0) line, which is assumed to be optically thick at all positions except diffuse cloud envelopes. We take $T_{ex}\sim12.5$ K for all clouds, based on typical peak $^{12}$CO(J=1--0) main beam temperatures in clouds A and B outside localized regions of active star formation. Cloud C $T_{mb}$ values are not included in this estimation, since its large distance and small angular extent mean that the filling factor of emission within the beam is likely to be low. 

The molecular mass traced in each line is obtained from the sum of $N_{\mathrm{H}_2}$ over all spatial positions at which emission is detected. Detection limits are derived on a cloud-by-cloud basis from the rms scatter in emission-free regions of the column density maps. In practice this means that cloud boundaries are defined at different absolute emission levels for each object. While less than ideal, this data-driven approach is unavoidable given the different noise and intensity levels in the three regions. 

In mass estimation we assume a distance of 2.6 kpc for clouds A and B and 6.5 kpc for cloud C, and include a factor of 1.4 to account for the presence of helium and heavier elements. Masses, peak column densities, and detection limits are summarized in table \ref{table1}, along with cloud sizes and peak main beam temperatures.

\subsection{Cloud A: Active Star Formation \& Translucent Molecular Gas Embedded in the Shell Walls}

\subsubsection{Morphology and Physical Properties}

\begin{figure}
\epsscale{1.0}
\plotone{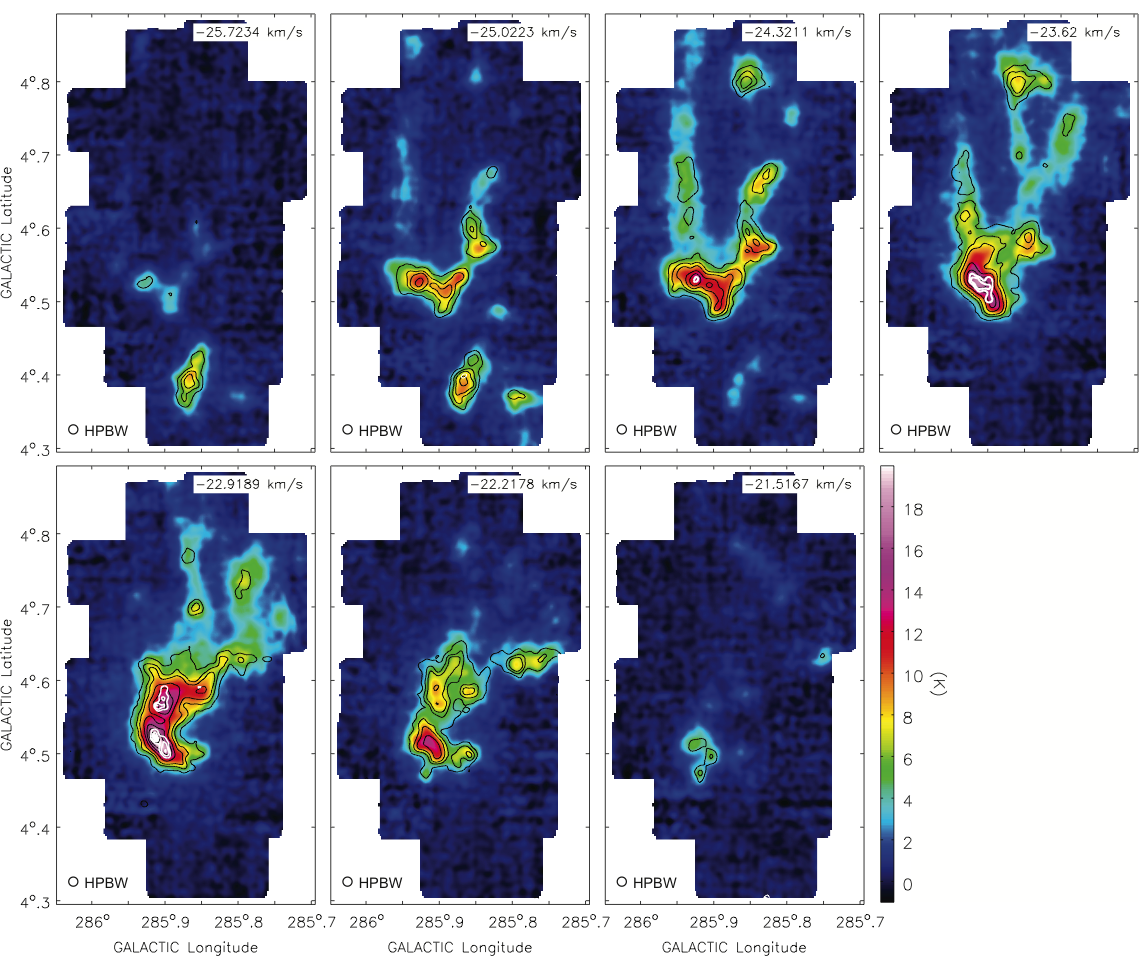}
\caption{Velocity channel maps of the embedded cloud, Cloud A. Each panel shows an average over four channels, corresponding to an interval of 0.7 km s$^{-1}$. The color scale shows $^{12}$CO(J=1--0) emission, black contours show $^{13}$CO(J=1--0) and white contours show C$^{18}$O(J=1--0). Contour levels start at the $4\sigma$ detection limit for each dataset, corresponding to 0.60 and 0.45 K for $^{13}$CO and C$^{18}$O respectively. Contour intervals in $^{13}$CO are every 0.5 K until 1.6K and every 1.0 K thereafter. Contour intervals in C$^{18}$O are every 0.1 K. C$^{18}$O contours are only plotted within the $^{12}$CO(J=1--0) cloud boundary (i.e. noise peaks from high-noise regions outside the cloud have been removed).}
\label{fig:cloudachs}
\end{figure}

Cloud A is the brightest and most massive of the three objects, with estimated masses of $4.9\times10^3~M_{\odot}$, $2.0\times10^3~M_{\odot}$ and $460~M_{\odot}$ in the regions traced by the $^{12}$CO, $^{13}$CO and C$^{18}$O lines, respectively. Figure \ref{fig:cloudachs} shows velocity channel maps of all three tracers, revealing much newly-resolved structure that was not visible in earlier low-resolution maps. 

The cloud is dominated by a bright nucleus that stretches from around $4.5^{\circ}$ to $4.6^{\circ}$ in Galactic latitude. The material in this region is traced in both $^{12}$CO(J=1--0) and $^{13}$CO(J=1--0), indicating relatively dense and high column density gas. The two $^{13}$CO emission peaks at $(l, b)=(285.91^\circ, 4.52^\circ)$ and $(285.90^\circ, 4.57^\circ)$ each host sizable C$^{18}$O cores, with linear extents of $\sim2$--3 pc and estimated molecular masses of $350~M_{\odot}$ and $110~M_{\odot}$ respectively. Peak column densities in both cores are $N_{\mathrm{H}_2}\sim8\times10^{21}$ cm$^{-2}$, and number densities are estimated to be $n\sim1500$. The presence of so much C$^{18}$O implies dense, possibly star-forming molecular gas, and indeed this entire central zone is associated with multiple indicators of active star formation, as will be discussed below.  

Above this dense nucleus, from latitudes of $b\approx4.63^{\circ}$ to the limits of the map at $b\approx4.88^{\circ}$, lie several fainter filamentary features. These extended, wispy structures are only partially detected in $^{13}$CO(J=1--0), indicating relatively diffuse, lower column density material. With angular widths of $\sim1\arcmin$--$4\arcmin$ ($\sim1$--$3$ pc), and orientations parallel to the main shell wall, these features are strongly reminiscent of substructure observed at higher latitudes in this section of wall in \citetalias{dawson11}. There we noted the presence of thin, twisted filaments between $4.9^{\circ} < b < 6.2^{\circ}$, with widths as narrow as the $\sim2.5\arcmin$ resolution limit of the data. These filaments were detected in H{\sc i} and also traced partially in CO, highlighting the tight relationship between the molecular and atomic material. The present data demonstrate that similar structural characteristics persist well into the most molecule-rich regions of the wall. In a scenario where Cloud A has formed in-situ from the swept-up medium, these diffuse molecular filaments are convincingly interpreted as material that is less advanced along an evolutionary pathway that runs from atomic to molecular to star-forming gas.

\subsubsection{Star Formation}

The bright core of Cloud A is actively forming high mass stars. It is home to the 2MASS stellar cluster DBSB 49 \citep{dutra03} -- a small open cluster with an estimated age of $2.1\pm0.3$ Myr, whose most massive member is of spectral type B0V \citep{soares08}. Two bright IRAS embedded young stellar object (YSO) candidates, IRAS 10489-5403 and IRAS 10492-5403, also lie within 
this core region, providing further evidence that star formation is still ongoing. Here, the IRAS source selection criteria are a `high' or `moderate' data quality flag at 25 $\mu$m and 60 $\mu$m, and a 60/25 $\mu$m flux density ratio of $>1.0$ \citep[see][]{beichman88}. Assuming a spectral energy distribution that peaks at 100 $\mu$m \citep{myers87}, the bolometric luminosities of these sources are estimated to be $1200~L_{\odot}$ and $1700~L_{\odot}$ -- equivalent to ZAMS spectral types of $\sim$B4--B5 if each source corresponds to a single object. The brighter of the two, IRAS 10492-5403, resides in the highest column density region of the lower C$^{18}$O core

\begin{figure}
\epsscale{1.0}
\plotone{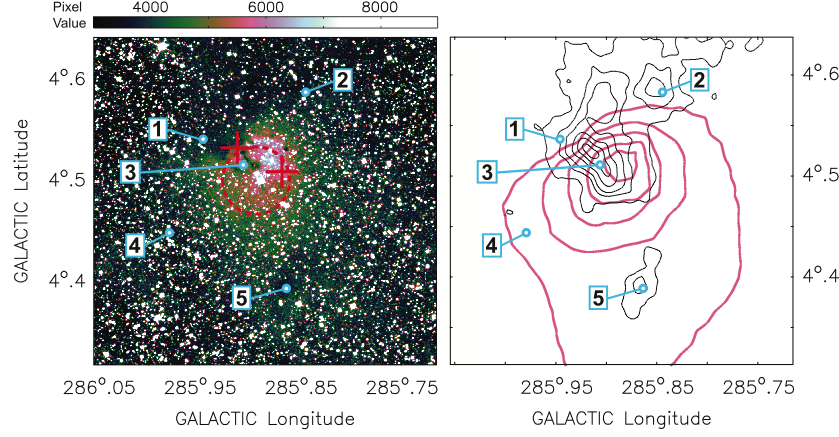}
\caption{Anatomy of the star forming region in Cloud A. The color image shows optical data from the second generation Digital Sky Survey (DSS2). The dashed circle marks the radius of the young IR cluster DBSB 49, while red crosses mark IRAS point sources with YSO colors. 
The contour image shows velocity-integrated $^{13}$CO(J=1--0) emission (black contours), scaled to units of H$_2$ column density, with levels at [0.7, 1.7, 2.7, 4.7, 6.7, 8.7]$\times10^{21}$ cm$^{-2}$. Pink contours are SHASSA H$\alpha$ emission beginning at a level of 55 Rayleigh (R) and incremented every 20 R thereafter. Numbered blue circles mark the locations of the spectra shown in Figure \ref{fig:specs}.}
\label{fig:sfrfig}
\end{figure}

The cluster is powering an optical H{\sc ii} region $\sim15\arcmin$ ($\sim11$ pc) in diameter. This H{\sc ii} region is seen in SHASSA H$\alpha$ emission maps \citep{gaustad01} as a bright compact source over an order of magnitude brighter than the weakly ionized inner rim of the shell walls \citep[see][]{dawson08a}. Figure \ref{fig:sfrfig} shows the red DSS2 optical image of this region, overlaid with the positions of the stellar cluster and IRAS sources described above, with $^{13}$CO and H$\alpha$ contours shown on an adjoining panel. The lowest H$\alpha$ contour marks the estimated boundary of the H{\sc ii} region, and is set at the level at which SHASSA H$\alpha$ emission drops to approximately $1.2$ times that of the local background. This boundary extends far beyond the edges of the CO emission, indicating that the H{\sc ii} region is no longer confined by its parent molecular cloud. Prominent dark dust lanes at $(l, b)\approx(285.91^{\circ}, 4.52^{\circ})$ coincide with the highest column-density regions of the molecular gas, and the small molecular clump at $(285.87^{\circ}, 4.39^{\circ})$ is also seen in absorption against the optical nebula.

The ionizing flux required to power the H{\sc ii} region provides an independent probe of the stellar population of the cluster. The expression for the radius of a classical Str\"{o}mgren sphere, $R=(3N^*/4\pi\alpha n_e^2)^{1/3}$, may be used to obtain a first order estimate of this flux under the assumption that ionization equilibrium has been reached. We estimate the electron density, $n_e$, from the emission measure via the relation, $EM$ (cm$^{-6}$ pc) $=2.75~T_4^{0.9}~I_{\mathrm{H}\alpha}~e^{2.2E(B-V)}=n_e^2L$, where $T_4$ is the temperature in $10^4$ K, $I_{\mathrm{H}\alpha}$ is the H$\alpha$ intensity in Rayleighs, $E(B-V)$ is the reddening due to material in front of the H{\sc ii} region, and $L$ is the linear depth of the emitting column. The reddening is obtained from the neutral gas column density via the relation $E(B-V)=A_V/3.1=N_{\mathrm{H}}/(5.9\times10^{21}~\mathrm{cm}^{-2})$ \citep[e.g.][]{draine96}. For the peak H$\alpha$ intensity of 140 R, a linear size of $L=11$ pc, and a typical temperature of $10^4$ K, we obtain $n_e\sim6$--$15$ cm$^{-3}$, where the lower value arises from ignoring the reddening term completely, and the higher from assuming the full column of molecular material at the peak position is located in front of the ionized gas. Taking the radius of the H{\sc ii} region as 5.5 pc, $N^*=[0.22$--$1.37]\times10^{48}$ s$^{-1}$. This value can be comfortably provided by the brightest cluster member, the B0V star \citep{vacca96}. We therefore find no need for additional massive stars hidden within the cloud.

Nevertheless, the level of star formation in Cloud A is highly unusual for such a small, high altitude molecular cloud. The scale height of massive star forming regions in the Galactic Disk is $\sim30$ pc, very similar to the scale height of main sequence OB stars \citep{bronfman00, urquhart11}. While massive star formation at higher altitudes is not unheard of \citep[e.g.][]{snell02}, at $z\sim200$ pc the star forming nucleus of Cloud A is still an extreme outlier in this distribution. In addition, the cloud mass is somewhat lower than typically observed for objects hosting B0 stars \citep{larson82,dobashi01}, suggesting a high star formation efficiency. The unambiguous identification of Cloud A as part of an expanding superstructure highlights the likely importance of supershells in providing star-forming molecular gas to the upper regions of the Galactic Disk.

\subsubsection{Dynamics}

\begin{figure}
\epsscale{1.0}
\plotone{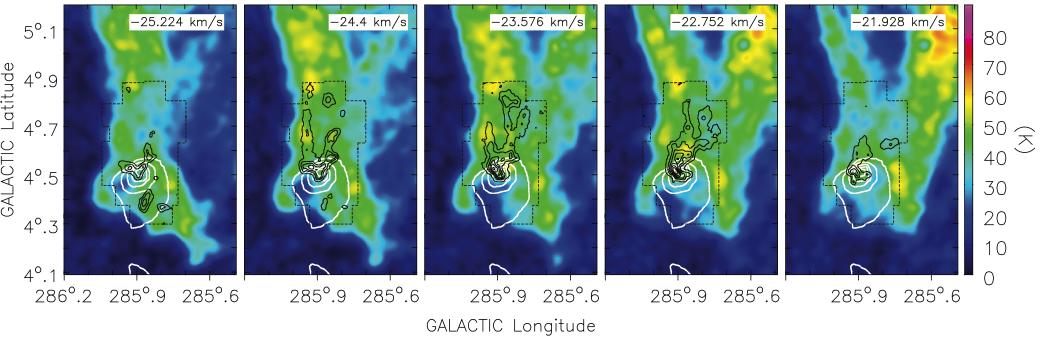}
\caption{Velocity channel maps of H{\sc i} and $^{12}$CO(J=1--0) in the vicinity of Cloud A. Color panels are H{\sc i}. Black contours are $^{12}$CO(J=1--0), beginning at 2.5 K and incremented every 3.0 K. White contours show SHASSA H$\alpha$ emission (with no velocity information) beginning at a level of 55 Rayleigh (R) and incremented every 20 R thereafter. The black dashed line marks the limits of the region observed in CO.}
\label{fig:cloudahihii}
\end{figure}

\begin{figure}
\epsscale{1.0}
\plotone{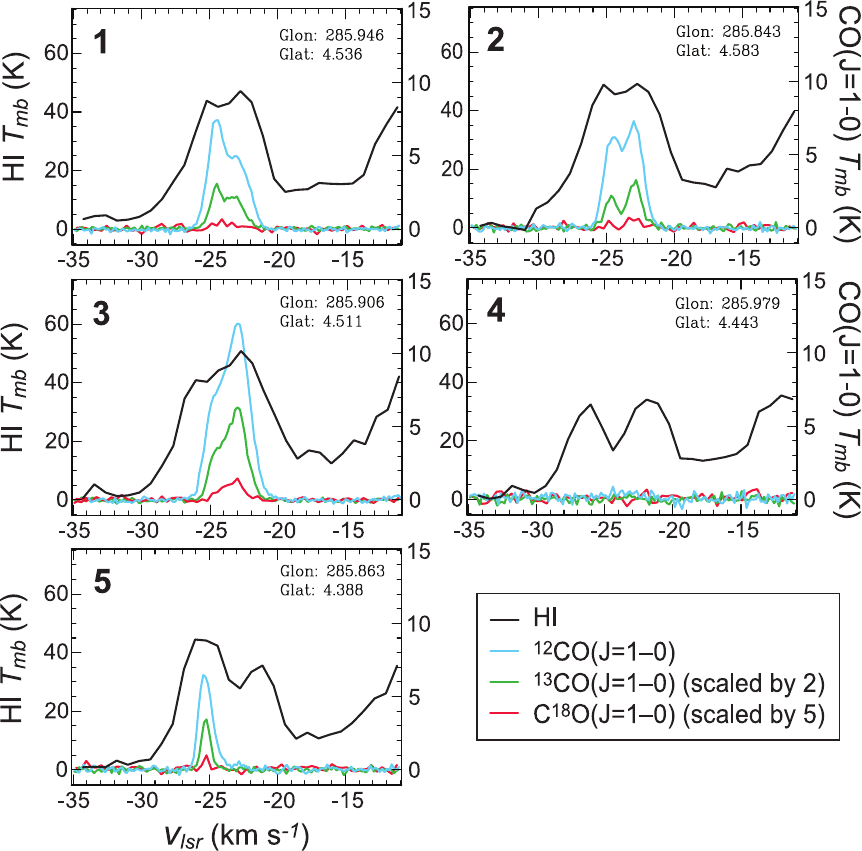}
\caption{H{\sc i} and CO(J=1--0) spectra at the positions numbered in Figure \ref{fig:sfrfig}. H{\sc i} data are from \citet{dawson11} and CO data have been smoothed to match the $2.5\arcmin$ resolution of the H{\sc i} cube. The left- and right-hand scales show H{\sc i} and CO intensities, respectively. $^{13}$CO(J=1--0) and C$^{18}$O(J=1--0) spectra have been scaled by factors of 2 and 5 respectively for ease of viewing.}
\label{fig:specs}
\end{figure}

The star formation activity strongly influences the dynamics of Cloud A and the region surrounding it. The molecular gas in the bright nucleus shows wide or multi-component spectral profiles, with typical velocity dispersions of $\Delta v\approx2.35\sigma_{v}\sim2$--3 km s$^{-1}$ in $^{12}$CO(J=1--0); significantly broader than elsewhere in the cloud. This dynamical influence is even more pronounced in the structure of the atomic material in the shell wall. Figure \ref{fig:cloudahihii} shows H{\sc i} channel maps overlaid with $^{12}$CO(J=1--0) and SHASSA H$\alpha$ contours. 
The H{\sc ii} region has carved out a cavity in the neutral gas, about which the H{\sc i} shows the classic signature of expansion -- a wide void at $v_{lsr}\approx-23$ km s$^{-1}$ that fills in at the velocity channels to either side. This behavior manifests in the H{\sc i} spectra as a characteristic broadened or double-peaked profiles, some examples of which are shown in Figure \ref{fig:specs}. 
This cavity opens out to the shell interior at its lower extreme, indicating that the H{\sc ii} region has not only broken out of its parent molecular cloud, but is also in the process of blistering out of the atomic shell wall in which the whole complex resides. 

Because the presence in H{\sc i} spectra of two peaks flanking a trough is also a characteristic signature of absorption -- either against a continuum background source or H{\sc i} self absorption (HISA) -- we take a moment to examine these two alternative hypotheses. In the case of continuum absorption, the 1.4 GHz continuum brightness of the H{\sc ii} region, as derived from the continuum data present in the original H{\sc i} observations, is only $\lesssim2$ K -- too low by an order of magnitude to explain the observed depths of the troughs. In the case of HISA, even when a clear trough is present, the outer slopes of the lines are too steep to be fitted by a single positive warm background component and negative cold foreground component, and are instead well fitted by two positive Gaussians \citepalias[see also][]{dawson11}. Most tellingly, distinct double-peaked or multi-component spectra are observed in all three CO lines over the same region. Unlike H{\sc i} these profile shapes cannot arise as a result of absorption, and must indicate the presence of separate velocity components. Taken together with the spatio-velocity structure of the neutral gas described above, the evidence therefore strongly favors a dynamical origin for the split profiles observed throughout this region. 

Despite a strong correlation between the spatial positions of multi-component H{\sc i} and CO spectra, the relationship between the emission components in the two tracers is generally not simple. Although both H{\sc i} and CO may exhibit multiple velocity components at any given position, these components do not in general correspond exactly, indicating some independence of movement between the atomic and molecular material, and more complex dynamics than a simple expanding shell. In particular, the velocity separation between CO components is typically no greater than $\sim2.5$ km s$^{-1}$, whereas for H{\sc i} this value reaches $\sim5.5$ km s$^{-1}$, with the largest separations observed in regions where CO is not detected. The origin of this behavior may lie in the different densities of the molecular and atomic components, the former of which is less easily accelerated by the energy input from the star forming region. 

The energy input that drives the expanding motion of the ISM around massive stars is provided both by over pressurization of the warm H{\sc ii} region, and by the mechanical energy input of stellar winds. Combined H{\sc ii} region and stellar wind models predict that $\sim0.1\%$ of the combined energy input into the system in the form of H-ionizing photons and wind luminosity is converted to kinetic energy of the surrounding gas over the lifetime of the star \citep{freyer03,freyer06}. Assuming separate expansion velocities of 2.75 km s$^{-1}$ and 1.25 km s$^{-1}$ for the atomic and molecular components, and associated masses of $\sim900~M_{\odot}$ and $\sim3.0\times10^3~M_{\odot}$, respectively, the kinetic energy of the neutral gas is estimated to be $\sim1.1\times10^{47}$ ergs. Here, the atomic mass is estimated from the fitted H{\sc i} spectra of \citetalias{dawson11}, assuming an optically thin medium, and the molecular mass is derived from $^{12}$CO(J=1--0) emission, with both tracers summed over the vicinity of the H$\alpha$ nebula. A B0V star emits Lyman continuum radiation at a rate of $\sim1.4\times10^{48}$ photons s$^{-1}$ \citep{vacca96}, which gives a lower limit of $\sim2.0\times10^{51}$ ergs for the total energy input in the form of ionizing radiation over the 2 Myr lifetime of the cluster, assuming a wavelength of 91.2 nm for all photons. The contribution from the stellar wind is considerably lower. The mass loss rates and terminal velocities of \citet{schaerer97} imply a total mechanical energy input of only $\sim7.6\times10^{48}$ ergs for a B0V star over the cluster lifetime. The total energy input in the form of Lyman continuum photons and mechanical wind luminosity is therefore taken to be $\sim2.0\times10^{51}$ ergs. The measured kinetic energy of the neutral gas is only $\sim0.01\%$ of this figure, implying that the energy input of the star is at least an order of magnitude greater than that required to drive the motion of the surrounding ISM. This discrepancy may arise partially from an underestimate of the associated neutral gas mass. Over half of the material in this section of wall is known to be either cold H{\sc i} or CO-dark H$_2$, and is hence not recovered by the standard mass estimation methods employed here \citepalias[see][]{dawson11}. Secondly, the fraction of input energy lost to the system by dust heating and cooling may be significant, and is not accounted for in the models of \citet{freyer03,freyer06}. Moreover, unlike the idealized model case, the H{\sc ii} region has broken out of its confining section of shell wall, allowing the escape of energy into the shell interior. Taking these factors into consideration, we consider the energy input of the central star to provide a reasonable explanation for the dynamics of the gas of the star-forming region. 

Outside the influence of the star forming region, above about $b\approx4.65^{\circ}$, CO velocity dispersions are smaller ($\Delta v\approx2.35\sigma_{v}\sim1$--2 km s$^{-1}$ in the $^{12}$CO line) and spectral profiles simpler. In cases where double components are present, they can generally be assigned to discrete spatial features in a way that is not possible in the more confused emission of the star-forming zone. H{\sc i} spectra are also generally well fit by single Gaussian profiles, although the presence of some barely resolved double-component lines \citepalias[see][]{dawson11} hints at further complexity that might be more clearly revealed by data at higher spatial or spectral resolutions. Nevertheless, there is no evidence to suggest the kind of deterministic motions observed in the lower regions of the cloud, and velocity structure here is well explained as a combination of turbulent and bulk motions imprinted on the wall during its formation. This difference between the dynamics of the star-forming nucleus and the more diffuse material outside it highlights a transition between a zone in which the properties of the ISM are governed by the physical conditions in the superstructure of which it is part, and the stage at which the local influence of star formation starts to dominate. Current estimates of the ages of GSH 287+04Ð17 ($\sim10^7$ yr) and the cluster ($2.1\pm0.3\times 10^6$ yr) imply that this transition began relatively recently in the shell's lifetime.


\subsection{Clouds B \& C: Dynamical Disruption of Pre-existing Molecular Gas \& Long-Lived Remnant Clumps}

\subsubsection{Morphology and Physical Properties}

Cloud B is smaller and less massive than Cloud A, with estimated masses of $830~M_{\odot}$, $310~M_{\odot}$ and $80~M_{\odot}$ traced in $^{12}$CO(J=1--0), $^{13}$CO(J=1--0) and C$^{18}$O(J=1--0), respectively. However, the fraction of the total mass traced by the rarer $^{13}$CO and C$^{18}$O lines is the same for both objects, at approximately $40\%$ and $10\%$. 

Cloud C is the smallest of the three clouds, with estimated masses $350~M_{\odot}$ and $70~M_{\odot}$ traced in $^{12}$CO(J=1--0) and $^{13}$CO(J=1--0), respectively. The mass fraction traced by the $^{13}$CO line is thus approximately half that of the other two clouds. No C$^{18}$O(J=1--0) is detected, to an estimated H$_2$ column density detection limit of $1.2\times10^{21}$ cm$^{-2}$. This translates to a mass detection limit of $\sim40~M_{\odot}$, meaning that a C$^{18}$O core with a fractional mass of $\sim10\%$, as seen in Clouds A and B, would not be detected in the present data.

\begin{figure}
\epsscale{1.0}
\plotone{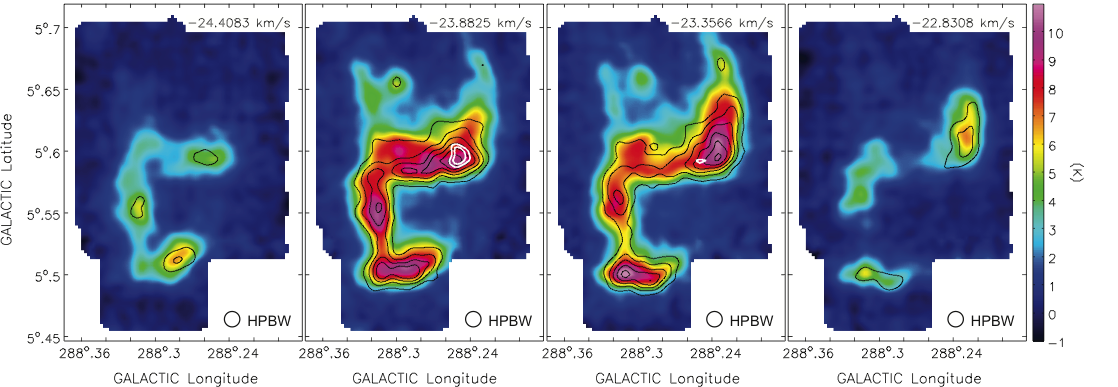}
\caption{Velocity channel maps of the offset cloud, Cloud B. Each panel shows an average over three channels, corresponding to an interval of 0.5 km s$^{-1}$. The color scale shows $^{12}$CO(J=1-0) emission, black contours show $^{13}$CO(J=1-0) and white contours show C$^{18}$O(J=1-0). Contour levels start at the $4\sigma$ detection limit for each dataset, corresponding to 0.7 and 0.4 K for $^{13}$CO and C$^{18}$O respectively. Contour intervals are every 0.5 K in $^{13}$CO and every 0.1 K in C$^{18}$O.}
\label{fig:cloudbchs}
\end{figure}

\begin{figure}
\epsscale{1.0}
\plotone{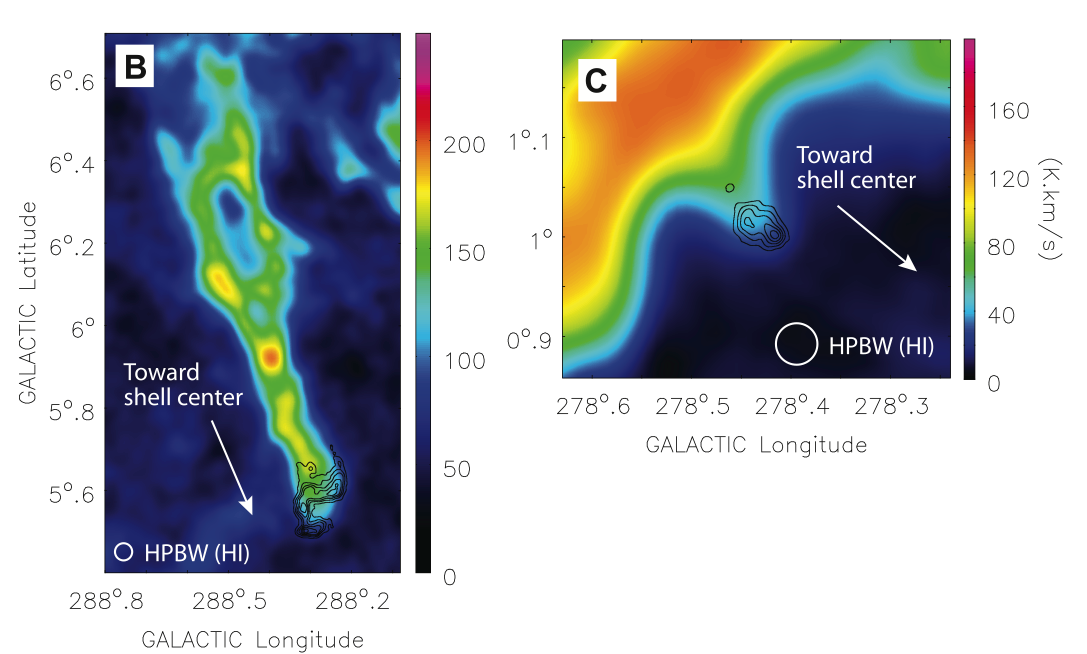}
\caption{H{\sc i} and $^{12}$CO(J=1--0) integrated intensity in the vicinity of Clouds B and C. Color is H{\sc i} and black contours are $^{12}$CO(J=1--0). For the Cloud B region, emission is integrated over the velocity range $-25.2 < v_{lsr} < -21.9$ km s$^{-1}$, and CO contours begin at 3.0 K km s$^{-1}$ and are incremented every 2.5 K km s$^{-1}$. For the Cloud C region, emission is integrated over the velocity range $41.2 < v_{lsr} < 43.7$ km s$^{-1}$, and CO contours begin at 2.0 K km s$^{-1}$ and are incremented every 1.0 K km s$^{-1}$.}
\label{fig:cloudbchi}
\end{figure}

Figures \ref{fig:cloudbchs} and \ref{fig:cloudbchi} show $^{12}$CO(J=1--0), $^{13}$CO(J=1--0) and C$^{18}$O(J=1--0) velocity channel maps of Cloud B, and the relationship of this molecular gas to the surrounding and atomic material. The data reveal new morphological evidence to support the interpretation that this dense-tipped head-tail feature was formed by the Between $(l, b)\approx(288.30^{\circ}, 5.58^{\circ})$ and $(288.24^{\circ}, 5.69^{\circ})$ the cloud shows a characteristic bow-like shape, with a sharp intensity gradient along the curved front edge facing the centre of the shell. This sharp edge is seen in both $^{12}$CO(J=1--0) and $^{13}$CO(J=1--0), with the latter concentrated towards the front of the feature, suggesting the preferential build up of denser gas there (see also Figure \ref{fig:nh2all}). This bow-shaped zone also hosts a C$^{18}$O core with an estimated mass of $\sim80~M_{\odot}$ and density of $n\sim1500$ cm$^{-3}$, located at the position of peak $^{13}$CO(J=1--0) intensity. To the right, the edge of the molecular cloud neatly hugs the border of the H{\sc i} emission, tapering to a narrow point at $(l, b)=(288.24^{\circ}, 5.69^{\circ})$. Behind this front, trailing fragments are seen at at $(l, b)\approx(288.31^{\circ}, 5.65^{\circ})$, following the direction of the H{\sc i} tail. These features are predominantly detected in $^{12}$CO alone, and appear to represent more diffuse molecular gas breaking away from the back edge of the cloud. 

In contrast, a bright $^{13}$CO-rich clump is seen lagging behind the expansion at $(l, b)\approx(288.31^{\circ}, 5.50^{\circ})$. At the present resolution of the H{\sc i} data, the peak of this molecular clump lies on the very edge of the atomic gas feature and may represent a particularly dense fragment of the disrupted cloud left behind as the surrounding envelope is stripped. The emission joining this clump to the rest of the cloud shows an unusually straight edge, and very square corners where it intersects the rest of the cloud. Surprisingly, this very regular squarish shape accurately reflects the global morphology of the gas. This can be readily verified by reprocessing the data from different scan directions separately, which demonstrates that the $90^{\circ}$ angle between the main components of the cloud is real. However, the extreme squareness is partially enhanced by the coincidental alignment of cloud structure with the telescope scan direction. 

\begin{figure}
\epsscale{1.0}
\plotone{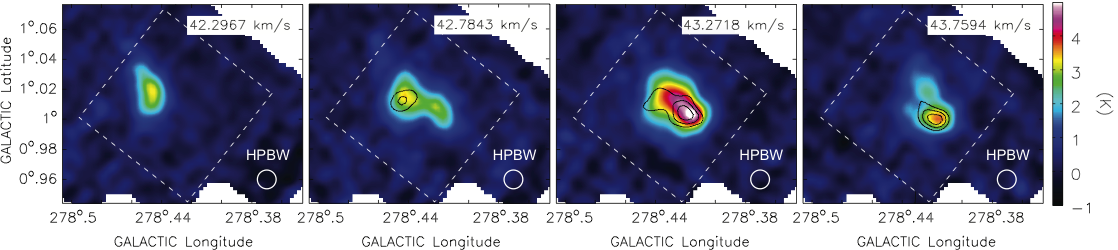}
\caption{Velocity channel maps of the offset cloud, Cloud C. Each panel shows an average over three channels, corresponding to an interval of 0.5 km s$^{-1}$. The color scale shows $^{12}$CO(J=1-0) emission and black contours show $^{13}$CO(J=1-0). Contour levels start at the $4\sigma$ detection limit of and 0.3 K and are incremented every 0.2 K. The white dashed lines mark the region observed in $^{13}$CO(J=1-0).}
\label{fig:cloudcchs}
\end{figure}

Cloud C, at a relatively large distance of 6.5 kpc, is resolved here for the first time, and shows many morphological similarities to its nearer cousin. Figures \ref{fig:cloudbchi} and \ref{fig:cloudcchs} show $^{12}$CO(J=1--0) and $^{13}$CO(J=1--0) velocity channel maps, and the relationship between this molecular material and the nearby atomic shell wall. The CO emission shows a distinct head-tail structure, with a bright peak located in both tracers at $(l, b)\approx(278.42^{\circ}, 1.00^{\circ})$, and weaker emission extended to the upper-left, pointing away from the centre of the shell. The head lies at the very tip of an atomic protrusion, about which the curved walls of the shell appear to have been carved out. As in the case of Cloud B, the edge facing the centre of the shell shows a sharper intensity gradient in both $^{12}$CO and $^{13}$CO than elsewhere in the cloud, with $^{13}$CO emission concentrated towards the leading edge. The molecular tail shows a second weaker emission peak at $(l, b)\approx(278.44^{\circ}, 1.02^{\circ})$, and beyond that a single drop of $^{12}$CO, trailing behind the main cloud at $(l, b)\approx(278.46^{\circ}, 1.05^{\circ})$. This barely detected feature is reminiscent of the trailing fragments observed in Cloud B. 

The wall of GSH 277+00+36 shows several examples of similar H{\sc i} drips, spaced periodically along the shell walls, with molecular matter so far detected at the tips of only two. Their origin has so far not been conclusively determined, and existing works have advanced the differing (although not necessarily mutually exclusive) interpretations of  these features as Rayleigh Taylor drips \citep{mcclure03} or as protrusions left behind through interaction of shell with pre-existing dense clumps \citepalias{dawson11}. The present observations provide further support for this latter interpretation.

\subsubsection{Cloud Dynamics and Clump Survival}

The spectral profiles of both CO and H{\sc i} in Clouds B and C are simple, and generally well fit by single Gaussian components at coincident velocities. $^{12}$CO(J=1--0) linewidths are typically $\Delta v_{\mathrm{FWHM}}\approx1.0$--1.5 km s$^{-1}$, and we see no direct kinematical evidence of cloud-shell interaction, such as shock-broadened line wings. This latter point is unsurprising. Based on the expansion velocities and projected distances of the molecular clouds from the main walls, the shell-cloud interaction would have occurred $>4$ Myr in the past for Cloud B and $>1$ Myr in the past for Cloud C. Assuming the velocity of the shock in the cloud material is given by $v_{cl}=v_{sh}~(\rho_{sh}/\rho_{cl})^{0.5}$ \citep[e.g.][]{klein94}, where $v_{sh}$ is the shell expansion velocity and $\rho_{sh}/\rho_{cl}$ is the mean shell-cloud density contrast, shocks will propagate completely through Clouds B and C on similar timescales of $\sim4$ Myr and $\sim1$ Myr, respectively, for shell wall densities of $n_{H}=10$ cm$^{-3}$ \citepalias{dawson11} and mean cloud densities of $n_{H}=600$ cm$^{-3}$ and $250$ cm$^{-3}$, as estimated from present $^{12}$CO data. 
This suggests that the shell shock fronts have now passed through both clouds, and initial shock signatures will have begun to dissipate. Any current dynamical interaction with the wider shell system is now presumably limited to the flow of the low-density interior medium past the remaining molecular gas; an interaction that is unlikely to produce strong spectral signatures. Moreover, shocked molecular gas is generally more readily traced in the higher excitation CO rotational lines, and may remain invisible in CO(J=1--0), even when present \citep[e.g.][]{reach05}. 

As pointed out in \citetalias{dawson11}, extinction arising from material outside the CO gas in Cloud B is insufficient to shield it against FUV ($6~\mathrm{eV}< h\nu<13.6~\mathrm{eV}$) photo-dissociation, and the transition between the molecular and atomic phases of the ISM is presumably proceeding in the molecular to atomic direction. However, while CO close to the surface of the cloud will be quickly destroyed, the high column densities at its peak positions imply long survival times for molecules deep within it. The CO destruction rate may be expressed as $I'_{\mathrm{CO}}G'_0f_{\mathrm{CO}}~e^{-2b_{\mathrm{CO}}A_V}$ s$^{-1}$ \citep[see Appendix C of][]{wolfire10}, where $I'_{\mathrm{CO}}$ is the unshielded CO photodissociation rate in the Draine field, $G'_0$ is the ratio of the incident radiation field to the Draine field, $f_{\mathrm{CO}}$ is the CO self shielding factor, and $b_{\mathrm{CO}}$ is the dust extinction coefficient for CO extinction. Estimating $A_V$ from $N_{\mathrm{H}_2}$ using the relation $A_V=N_{\mathrm{H}}/(1.9\times10^{21}~\mathrm{cm}^{-2})$, and taking all other parameters as in \citetalias{dawson11}, we obtain characteristic dissociation timescales of $\tau\gtrsim10^8$ yr for molecules at the centre of the bright peaks of the cloud. While this simplistic treatment does not account for the time-variation of the destruction rate with decreased self-shielding nor the ongoing recombination of CO molecules, it nevertheless suggests that the gas in its present state cannot be easily rendered unobservable by photodissociation, provided the background UV field is not substantially higher than in the solar neighborhood. This is a reasonable assumption given the low level of ionization observed around the inner rim of GSH 287+04-17 \citep{dawson08a}. 

Other destructive influences on the molecular gas include thermal evaporation from the hot interior medium and dynamical interaction with the shell system. The classical evaporation timescale for Cloud B was estimated to be $\sim10^8$ yr \citepalias{dawson11}, and this figure does not change substantially as a result of the resolution improvements of the present observations. In contrast, dynamical interaction with the shell is likely to initially disrupt the clouds on much shorter characteristic timescales of a few Myr \citepalias{dawson11}. However, the most intense phase of this interaction is limited to the finite period in which the initial shock imparted through interaction with the dense material of the shell is still propagating through the cloud; a stage that Cloud B has already survived. For a canonical interior medium density of $\sim0.01$ cm$^{-3}$ and the slow flow-speed associated with the expansion of the shell, the current dynamical influence on the clouds is minimal.

In the absence of a continuing disruptive influence from the shell system, the fate of Cloud B will be determined by the balance between its own internal velocity dispersion  and the confining forces of external pressure and gravity. Assuming a spherical cloud with a $r^{-1}$ radial density profile and ignoring external pressure and magnetic fields, the virial mass required for self-gravity to just balance the system is given by $M_{vir}/M_{\odot}=190~R~(\Delta v_{cld})^2$, where $R$ is in parsecs and $\Delta v_{cld}$ is the FWHM velocity dispersion in km s$^{-1}$. Under this formulation, although the cloud as a whole is marginally unbound, the virial mass of the C$^{18}$O core is just sufficient for it to be self-gravitating ($M_{vir}=80~M_{\odot}$).

Performing the same analysis for Cloud C, we may make an identical argument concerning the minimal destructive influence due to evaporation by the internal medium and the current dynamical interaction of the cloud with the shell system. However, the characteristic photodissociation timescale for CO at the centre of the cloud is found to be much shorter, at $\sim10^6$ yr. This timescale is very sensitive to beam dilution effects, however, and may increase significantly in better resolved data. In addition, H$\alpha$ data for this region is confused, precluding comment on the likely magnitude of background UV field, and rendering the analysis inconclusive. The virial mass of Cloud C is found to be several times larger than its CO intensity derived mass, suggesting that it will disperse if not confined by external pressure.

\subsubsection{Comments on Star Formation}

Self-gravitating C$^{18}$O cores are a good predictor of star-forming gas \citep{tachihara02}, and we might therefore hope to discover signs of ongoing star formation in Cloud B. However, although several IRAS sources fall within the cloud boundary, no clear evidence is found. A single IRAS source with YSO colors is located at $(l, b)=(288.29^{\circ}, 5.58^{\circ})$, but is not associated with any other known star formation tracers or major CO intensity peaks. Two sources that do not fulfill the YSO selection criteria are also located within the cloud boundaries at $(l, b)=(288.26^{\circ}, 5.59^{\circ})$ and $(288.30^{\circ}, 5.50^{\circ})$. These sources are detected at 100 $\mu$m and 60 $\mu$m and at 60 $\mu$m only, and are both located close to CO emission peaks. However, in the absence of any corroborating evidence, it is unclear whether any of these sources trace embedded star formation or point-source-scale structure in the cloud itself. Nevertheless, the low bolometric luminosities of all three ($L_{bol}\lesssim100~L_{\odot}$) allow us to constrain any star formation that may be occurring to low mass objects. 

\section{Summary \& Conclusions}
\label{summary}

The transition between atomic and molecular gas forms a key link in the lifecycle of material in galactic systems. In a dynamic and multi-phase ISM, large-scale stellar feedback can drive this transition in either direction. Diffuse material accumulated and compressed in supershells can reach sufficient densities and column densities for the formation of molecular gas. Conversely, pre-existing molecular clouds may be disrupted by their encounter with an expanding shell shock front, leaving them stripped of shielding material and vulnerable to UV dissociation. 

In this context we have performed an in-depth case-study of three molecular clouds in the walls of the Galactic supershells GSH 287+04--17 and GSH 277+00+36. These clouds have been identified in previous work as likely locations at which the atomic-molecular transition is proceeding in opposite directions. All three objects were mapped to a resolution of $\sim45\arcsec$ in the $^{12}$CO(J=1--0), $^{13}$CO(J=1--0) and C$^{18}$O(J=1--0) lines, providing detailed information on the distribution and properties of the molecular ISM, and enabling a greatly improved discussion of its relationship to the wider environment in which it resides.

Clouds B and C are small condensations of molecular gas with molecular masses of $\sim830~M_{\odot}$ and $\sim350~M_{\odot}$, respectively, whose location at the tips of finger-like extensions of H{\sc i} suggests the sweep-over and subsequent entraining of pre-existing clouds. Our observations reveal hitherto unseen morphological signatures of disruption and possible shock-interaction in both clouds, strongly supporting this interpretation. Both objects show dense heads with sharp intensity gradients at their leading edges, and trailing tails of more diffuse material. The emission of Cloud B, the better resolved of the two, closely delineates the curved front rim of the long H{\sc i} feature with which it is associated, with molecular gas forming a bow-like shape -- a characteristic signature of a shocked cloud. 

Both clouds have survived the original interaction with their respective shells, and are now interacting with the diffuse interior medium. High density contrasts between the molecular gas and this interior medium suggest that the destructive dynamical and thermal effects of the shell systems at the present epoch are minimal; i.e. while the interaction with an expanding shell may have carved the ISM into the configurations we currently observe, the shell systems are no longer a primary determinant of the cloud lifetimes. In the case of Cloud B, high peak column densities and an absence of nearby sources of FUV photons implies that although photodissociation of CO is underway, and the gas is becoming atomic, this process is proceeding at a relatively gentle pace, suggesting potentially long survival times if the cloud remains in its current configuration. However, the internal velocity dispersions of both Clouds B and C suggest that neither are globally self-gravitating, and may disperse if not confined by external pressure. Cloud B does, however, contain a dense C$^{18}$O core which is locally bound. Such cores are generally good predictors of star-forming gas, and the presence of several cool, low-luminosity IRAS sources within the cloud hints at possible low mass star formation activity.

Cloud A is a moderately sized, massive-star-forming cloud embedded within the main wall of GSH 287+04--17. With a substantial molecular mass of $\sim4.9\times10^3~M_{\odot}$, and a distribution that closely follows that of the curved atomic shell, it has been identified in previous work as a likely example of molecular material formed in-situ from the swept-up atomic ISM. Our observations reveal a bright star-forming nucleus detected in all three CO lines, from which emanate several diffuse filaments detected primarily in $^{12}$CO(J=1--0) alone.

The star formation in Cloud A is a convincing example of second generation massive star formation triggered by the primary generation that formed the shell. The star forming nucleus hosts the young ($2.1\pm0.3$ Myr) cluster DSBS 49, whose most massive member is of spectral type B0V. The formation of such massive stars is highly unusual at this Galactic altitude ($z\sim200$ pc), and the unambiguous identification of this region as part of an expanding superstructure highlights the role of supershells in providing star-forming material to the upper regions of the Galactic Disk. The cluster is powering an optical H{\sc ii} region $\sim11$ pc in diameter, which has carved out a cavity in the shell wall and is in the process of blistering out into the interior. Expansive motions driven by this star formation activity dominate the dynamics of both the atomic and molecular gas in this local region. 

In contrast, the filaments outside of the influence of this star-forming zone form quiescent parts of the shell wall. These diffuse, wispy-looking molecular features are $\sim1$--$3$ pc in width, and closely resemble twisted, filamentary structure detected in both H{\sc i} and CO elsewhere in the same section of wall (Paper 1), demonstrating that common structural characteristics persist throughout much of its volume. In a scenario where Cloud A has formed in-situ from the swept-up medium, these diffuse molecular filaments are interpreted as material that is less advanced along an evolutionary pathway that runs from atomic to molecular to star-forming gas. The border between these filaments and the star forming nucleus of the cloud marks a transition in which the properties of the ISM switch from being determined by the global characteristics of the swept-up material to being dominated by the local influence of the stellar sources forming within it. 
This critical stage in the process of triggered star/cloud formation is currently ongoing in the walls of GSH 287+04--17, $\sim10^7$ yr after its birth.

The data presented in this work mark all three clouds as excellent targets for further observation, setting the stage for an empirical study of the transition both ways between molecular and atomic gas.

\acknowledgements
The Australia Telescope Compact Array, Parkes and Mopra Telescopes are part of the Australia Telescope which is funded by the Commonwealth of Australia for operation as a National Facility managed by CSIRO. We acknowledge the use of the Southern H-Alpha Sky Survey Atlas (SHASSA), which is supported by the National Science Foundation. The DSS2 image used in this work is based on photographic data obtained using The UK Schmidt Telescope. The UK Schmidt Telescope was operated by the Royal Observatory Edinburgh, with funding from the UK Science and Engineering Research Council, until 1988 June, and thereafter by the Anglo-Australian Observatory. Original plate material is copyright of the Royal Observatory Edinburgh and the Anglo-Australian Observatory. The plates were processed into the present compressed digital form with their permission. The Digitized Sky Survey was produced at the Space Telescope Science Institute under US Government grant NAG W-2166.

\bibliographystyle{apj}
\bibliography{msbib}

\begin{thebibliography}{46}
\expandafter\ifx\csname natexlab\endcsname\relax\def\natexlab#1{#1}\fi

\bibitem[{{Audit} \& {Hennebelle}(2005)}]{audit05}
{Audit}, E., \& {Hennebelle}, P. 2005, \aap, 433, 1

\bibitem[{{Beichman} {et~al.}(1988){Beichman}, {Neugebauer}, {Habing}, {Clegg},
  \& {Chester}}]{beichman88}
{Beichman}, C.~A., {Neugebauer}, G., {Habing}, H.~J., {Clegg}, P.~E., \&
  {Chester}, T.~J. 1988, {Infrared astronomical satellite (IRAS) catalogs and
  atlases. Volume 1: Explanatory supplement}, Vol.~1 (Washington, DC: GPO)

\bibitem[{{Bergin} {et~al.}(2004){Bergin}, {Hartmann}, {Raymond}, \&
  {Ballesteros-Paredes}}]{bergin04}
{Bergin}, E.~A., {Hartmann}, L.~W., {Raymond}, J.~C., \& {Ballesteros-Paredes},
  J. 2004, \apj, 612, 921

\bibitem[{{Bronfman} {et~al.}(2000){Bronfman}, {Casassus}, {May}, \&
  {Nyman}}]{bronfman00}
{Bronfman}, L., {Casassus}, S., {May}, J., \& {Nyman}, L.-{\AA}. 2000, \aap,
  358, 521

\bibitem[{{Dawson} {et~al.}(2008{\natexlab{a}}){Dawson}, {Kawamura}, {Mizuno},
  {Onishi}, \& {Fukui}}]{dawson08b}
{Dawson}, J.~R., {Kawamura}, A., {Mizuno}, N., {Onishi}, T., \& {Fukui}, Y.
  2008{\natexlab{a}}, \pasj, 60, 1297

\bibitem[{{Dawson} {et~al.}(2011){Dawson}, {McClure-Griffiths}, {Kawamura},
  {Mizuno}, {Onishi}, {Mizuno}, \& {Fukui}}]{dawson11}
{Dawson}, J.~R., {McClure-Griffiths}, N.~M., {Kawamura}, A., et al. 2011, \apj, 728, 127

\bibitem[{{Dawson} {et~al.}(2008{\natexlab{b}}){Dawson}, {Mizuno}, {Onishi},
  {McClure-Griffiths}, \& {Fukui}}]{dawson08a}
{Dawson}, J.~R., {Mizuno}, N., {Onishi}, T., {McClure-Griffiths}, N.~M., \&
  {Fukui}, Y. 2008{\natexlab{b}}, \mnras, 387, 31

\bibitem[{{Dickman}(1978)}]{dickman78}
{Dickman}, R.~L. 1978, \apjs, 37, 407

\bibitem[{{Dobashi} {et~al.}(2001){Dobashi}, {Yonekura}, {Matsumoto}, {Momose},
  {Sato}, {Bernard}, \& {Ogawa}}]{dobashi01}
{Dobashi}, K., {Yonekura}, Y., {Matsumoto}, T., et al. 2001, \pasj, 53, 85

\bibitem[{{Draine} \& {Bertoldi}(1996)}]{draine96}
{Draine}, B.~T., \& {Bertoldi}, F. 1996, \apj, 468, 269

\bibitem[{{Dutra} {et~al.}(2003){Dutra}, {Bica}, {Soares}, \&
  {Barbuy}}]{dutra03}
{Dutra}, C.~M., {Bica}, E., {Soares}, J., \& {Barbuy}, B. 2003, \aap, 400, 533

\bibitem[{{Frerking} {et~al.}(1982){Frerking}, {Langer}, \&
  {Wilson}}]{frerking82}
{Frerking}, M.~A., {Langer}, W.~D., \& {Wilson}, R.~W. 1982, \apj, 262, 590

\bibitem[{{Freyer} {et~al.}(2003){Freyer}, {Hensler}, \& {Yorke}}]{freyer03}
{Freyer}, T., {Hensler}, G., \& {Yorke}, H.~W. 2003, \apj, 594, 888

\bibitem[{{Freyer} {et~al.}(2006){Freyer}, {Hensler}, \& {Yorke}}]{freyer06}
---. 2006, \apj, 638, 262

\bibitem[{{Fukui} {et~al.}(1999){Fukui}, {Onishi}, {Abe}, {Kawamura},
  {Tachihara}, {Yamaguchi}, {Mizuno}, \& {Ogawa}}]{fukui99}
{Fukui}, Y., {Onishi}, T., {Abe}, R., et al. 1999, \pasj, 51, 751

\bibitem[{{Gaustad} {et~al.}(2001){Gaustad}, {McCullough}, {Rosing}, \& {Van
  Buren}}]{gaustad01}
{Gaustad}, J.~E., {McCullough}, P.~R., {Rosing}, W., \& {Van Buren}, D. 2001,
  \pasp, 113, 1326

\bibitem[{{Hartmann} {et~al.}(2001){Hartmann}, {Ballesteros-Paredes}, \&
  {Bergin}}]{hartmann01}
{Hartmann}, L., {Ballesteros-Paredes}, J., \& {Bergin}, E.~A. 2001, \apj, 562,
  852

\bibitem[{{Heiles}(1979)}]{heiles79}
{Heiles}, C. 1979, \apj, 229, 533

\bibitem[{{Heitsch} \& {Hartmann}(2008)}]{heitsch08c}
{Heitsch}, F., \& {Hartmann}, L. 2008, \apj, 689, 290

\bibitem[{{Heitsch} {et~al.}(2008){Heitsch}, {Hartmann}, \&
  {Burkert}}]{heitsch08b}
{Heitsch}, F., {Hartmann}, L.~W., \& {Burkert}, A. 2008, \apj, 683, 786

\bibitem[{{Hennebelle} \& {P{\'e}rault}(1999)}]{hennebelle99}
{Hennebelle}, P., \& {P{\'e}rault}, M. 1999, \aap, 351, 309

\bibitem[{{Heyer} {et~al.}(2001){Heyer}, {Carpenter}, \& {Snell}}]{heyer01}
{Heyer}, M.~H., {Carpenter}, J.~M., \& {Snell}, R.~L. 2001, \apj, 551, 852

\bibitem[{{Hughes} {et~al.}(2010){Hughes}, {Wong}, {Ott}, {Muller}, {Pineda},
  {Mizuno}, {Bernard}, {Paradis}, {Maddison}, {Reach}, {Staveley-Smith},
  {Kawamura}, {Meixner}, {Kim}, {Onishi}, {Mizuno}, \& {Fukui}}]{hughes10}
{Hughes}, A., {et~al.} 2010, \mnras, 406, 2065

\bibitem[{{Hunter} {et~al.}(1997){Hunter}, {Bertsch}, {Catelli}, {Dame},
  {Digel}, {Dingus}, {Esposito}, {Fichtel}, {Hartman}, {Kanbach}, {Kniffen},
  {Lin}, {Mayer-Hasselwander}, {Michelson}, {von Montigny}, {Mukherjee},
  {Nolan}, {Schneid}, {Sreekumar}, {Thaddeus}, \& {Thompson}}]{hunter97}
{Hunter}, S.~D., {et~al.} 1997, \apj, 481, 205

\bibitem[{{Inoue} \& {Inutsuka}(2009)}]{inoue09}
{Inoue}, T., \& {Inutsuka}, S. 2009, \apj, 704, 161

\bibitem[{{Johansson} {et~al.}(1998){Johansson}, {Greve}, {Booth}, {Boulanger},
  {Garay}, {de Graauw}, {Israel}, {Kutner}, {Lequeux}, {Murphy}, {Nyman}, \&
  {Rubio}}]{johansson98}
{Johansson}, L.~E.~B., {et~al.} 1998, \aap, 331, 857

\bibitem[{{Klein} {et~al.}(1994){Klein}, {McKee}, \& {Colella}}]{klein94}
{Klein}, R.~I., {McKee}, C.~F., \& {Colella}, P. 1994, \apj, 420, 213

\bibitem[{{Koyama} \& {Inutsuka}(2000)}]{koyama00}
{Koyama}, H., \& {Inutsuka}, S. 2000, \apj, 532, 980

\bibitem[{{Ladd} {et~al.}(2005){Ladd}, {Purcell}, {Wong}, \&
  {Robertson}}]{ladd05}
{Ladd}, N., {Purcell}, C., {Wong}, T., \& {Robertson}, S. 2005, \pasa, 22, 62

\bibitem[{{Larson}(1982)}]{larson82}
{Larson}, R.~B. 1982, \mnras, 200, 159

\bibitem[{{Malhotra}(1994)}]{malhotra94}
{Malhotra}, S. 1994, \apj, 433, 687

\bibitem[{{McClure-Griffiths} {et~al.}(2002){McClure-Griffiths}, {Dickey},
  {Gaensler}, \& {Green}}]{mcclure02}
{McClure-Griffiths}, N.~M., {Dickey}, J.~M., {Gaensler}, B.~M., \& {Green},
  A.~J. 2002, \apj, 578, 176

\bibitem[{{McClure-Griffiths} {et~al.}(2003){McClure-Griffiths}, {Dickey},
  {Gaensler}, \& {Green}}]{mcclure03}
---. 2003, \apj, 594, 833

\bibitem[{{McClure-Griffiths} {et~al.}(2000){McClure-Griffiths}, {Dickey},
  {Gaensler}, {Green}, {Haynes}, \& {Wieringa}}]{mcclure00}
{McClure-Griffiths}, N.~M., {Dickey}, J.~M., {Gaensler}, B.~M., et al. 2000, \aj, 119, 2828

\bibitem[{{McCray} \& {Kafatos}(1987)}]{mccray87}
{McCray}, R., \& {Kafatos}, M. 1987, \apj, 317, 190

\bibitem[{{Myers} {et~al.}(1987){Myers}, {Fuller}, {Mathieu}, {Beichman},
  {Benson}, {Schild}, \& {Emerson}}]{myers87}
{Myers}, P.~C., {Fuller}, G.~A., {Mathieu}, R.~D., et al. 1987, \apj, 319, 340

\bibitem[{{Reach} {et~al.}(2005){Reach}, {Rho}, \& {Jarrett}}]{reach05}
{Reach}, W.~T., {Rho}, J., \& {Jarrett}, T.~H. 2005, \apj, 618, 297

\bibitem[{{Schaerer} \& {de Koter}(1997)}]{schaerer97}
{Schaerer}, D., \& {de Koter}, A. 1997, \aap, 322, 598

\bibitem[{{Snell} {et~al.}(2002){Snell}, {Carpenter}, \& {Heyer}}]{snell02}
{Snell}, R.~L., {Carpenter}, J.~M., \& {Heyer}, M.~H. 2002, \apj, 578, 229

\bibitem[{{Soares} {et~al.}(2008){Soares}, {Bica}, {Ahumada}, \&
  {Clari{\'a}}}]{soares08}
{Soares}, J.~B., {Bica}, E., {Ahumada}, A.~V., \& {Clari{\'a}}, J.~J. 2008,
  \aap, 478, 419

\bibitem[{{Tachihara} {et~al.}(2002){Tachihara}, {Onishi}, {Mizuno}, \&
  {Fukui}}]{tachihara02}
{Tachihara}, K., {Onishi}, T., {Mizuno}, A., \& {Fukui}, Y. 2002, \aap, 385,
  909

\bibitem[{{Urquhart} {et~al.}(2011){Urquhart}, {Moore}, {Hoare}, {Lumsden},
  {Oudmaijer}, {Rathborne}, {Mottram}, {Davies}, \& {Stead}}]{urquhart11}
{Urquhart}, J.~S., {et~al.} 2011, \mnras, 410, 1237

\bibitem[{{Vacca} {et~al.}(1996){Vacca}, {Garmany}, \& {Shull}}]{vacca96}
{Vacca}, W.~D., {Garmany}, C.~D., \& {Shull}, J.~M. 1996, \apj, 460, 914

\bibitem[{{V{\'a}zquez-Semadeni} {et~al.}(2007){V{\'a}zquez-Semadeni},
  {G{\'o}mez}, {Jappsen}, {Ballesteros-Paredes}, {Gonz{\'a}lez}, \&
  {Klessen}}]{vazquez07}
{V{\'a}zquez-Semadeni}, E., {G{\'o}mez}, G.~C., {Jappsen}, A.~K., et al. 2007,
  \apj, 657, 870

\bibitem[{{V{\'a}zquez-Semadeni} {et~al.}(2006){V{\'a}zquez-Semadeni}, {Ryu},
  {Passot}, {Gonz{\'a}lez}, \& {Gazol}}]{vazquez06}
{V{\'a}zquez-Semadeni}, E., {Ryu}, D., {Passot}, T., {Gonz{\'a}lez}, R.~F., \&
  {Gazol}, A. 2006, \apj, 643, 245

\bibitem[{{Wolfire} {et~al.}(2010){Wolfire}, {Hollenbach}, \&
  {McKee}}]{wolfire10}
{Wolfire}, M.~G., {Hollenbach}, D., \& {McKee}, C.~F. 2010, \apj, 716, 1191

\end{thebibliography}

\begin{deluxetable}{ccccccccccccccccccc}
\tabletypesize{\scriptsize}
\setlength{\tabcolsep}{0.04in} 
\tablecolumns{19} 
\tablewidth{0pc} 
\tablecaption{Summary of Cloud Properties} 
\tablehead{ 
\colhead{} & \colhead{} & \multicolumn{5}{c}{$^{12}$CO(J=1--0)} &  \colhead{} & \multicolumn{5}{c}{$^{13}$CO(J=1--0)} & \colhead{} & \multicolumn{5}{c}{C$^{18}$O(J=1--0)}\\ 
\cline{3-7} \cline{9-13} \cline{15-19} \\ 
\colhead{Cloud} & \colhead{} & \colhead{$T_{max}$\tablenotemark{a}} &\colhead{$N_{\mathrm{H}_2,lim}$\tablenotemark{b}} & \colhead{$N_{\mathrm{H}_2,max}$\tablenotemark{c}} & \colhead{$A_{cld}$\tablenotemark{d}} & \colhead{$M_{cld}$\tablenotemark{e}} & \colhead{} & \colhead{$T_{max}$\tablenotemark{a}} &\colhead{$N_{\mathrm{H}_2,lim}$\tablenotemark{b}} & \colhead{$N_{\mathrm{H}_2,max}$\tablenotemark{c}} & \colhead{$A_{cld}$\tablenotemark{d}} & \colhead{$M_{cld}$\tablenotemark{e}} & \colhead{} & \colhead{$T_{max}$\tablenotemark{a}} &\colhead{$N_{\mathrm{H}_2,lim}$\tablenotemark{b}} & \colhead{$N_{\mathrm{H}_2,max}$\tablenotemark{c}} & \colhead{$A_{cld}$\tablenotemark{d}} & \colhead{$M_{cld}$\tablenotemark{e}}\\
\colhead{} & \colhead{} & \colhead{K} &\colhead{$10^{21}$cm$^{-2}$} & \colhead{$10^{21}$cm$^{-2}$} & \colhead{pc$^2$} & \colhead{$M_{\odot}$} & \colhead{} & \colhead{K} &\colhead{$10^{21}$cm$^{-2}$} & \colhead{$10^{21}$cm$^{-2}$} & \colhead{pc$^2$}& \colhead{$M_{\odot}$} & \colhead{} & \colhead{K} &\colhead{$10^{21}$cm$^{-2}$} & \colhead{$10^{21}$cm$^{-2}$} & \colhead{pc$^2$}& \colhead{$M_{\odot}$}    
}
\startdata
A & & 21.8 & 0.56 & 9.0 & 155 & 4900 & & 7.4 & 0.55 & 9.9 & 57 & 1950 & & 1.1 & 3.6 & 8.5 & 4.8 & 460 \\
B & & 12.5 & 0.32 & 2.7 & 43 & 830 & & 4.3 & 0.40 & 2.2 & 18 & 310 & & 1.1 & 2.4 & 5.1 & 1.1 & 80 \\
C & & 5.4 & 0.24 & 1.1 & 39 & 350 & & 1.0 & 0.15 & 0.46 & 18 & 70 & & $<0.2$ & 1.2 & \nodata & \nodata & $<40$ \\
\enddata
\tablenotetext{a}{Peak main beam temperature}
\tablenotetext{b}{$3\sigma$ detection limit in H$_2$ column density.} 
\tablenotetext{c}{Peak value of H$_2$ column density.}
\tablenotetext{d}{Combined area of all pixels above H$_2$ column density detection limit.}
\tablenotetext{e}{Total molecular gas mass traced in each CO line. Includes factor of 1.4 to account for presence of He and heavier elements.}
\label{table1}
\end{deluxetable}

\end{document}